\documentclass[]{jpsj2}
\usepackage{graphicx}
\usepackage{latexsym}
\usepackage{amsmath}
\usepackage{bm}

\title{%
Fluctuation Induced Homochirality
}

\author{%
Takeshi \textsc{Sugimori}\thanks{tsugimor@rk.phys.keio.ac.jp}, 
Hiroyuki \textsc{Hyuga}\thanks{hyuga@rk.phys.keio.ac.jp},
and Yukio \textsc{Saito}\thanks{yukio@rk.phys.keio.ac.jp}
}

\inst{%
Department of Physics, Keio University, Yokohama 223-8522 
}

\recdate{\today}

\abst{
We propose a new mechanism 
for the achievment of homochirality in life 
without any autocatalytic production process. 
Our model consists of a spontaneous production together 
 with a recycling cross inhibition in a closed system. 
 It is shown that although
 the  rate equations for this system predict no chiral symmetry breaking,  
 the stochastic master equation predicts complete homochirality. 
 This is because the fluctuation induced by the discreteness  of population numbers of participating molecules plays essential roles. This fluctuation conspires with the recyling cross inhibition to realize the homochirality.

 }

\kword{%
homochirality,
probability distribution, 
master equation,
recycling cross inhibition, 
directed random walk
}

\begin{document}
\sloppy
\maketitle

\section{Introduction and Model System}
It has long been known since the discovery of Pasteur 
that organic molecules in life are homochiral, in other words, 
having a completely broken chiral symmetry\cite{pasteur849,japp898}.
The origin of this homochirality remains an unsolved important puzzle
\cite{stryer98,calvin69,goldanskii+88,gridnev06}.
Various mechanisms for the germination of chirality imbalance have been proposed such as 
different intensities of circularly polarized light in a primordial era, adsorption on optically active crystals, or the parity breaking in the weak interaction. 
\cite{feringa+99}

Predicted asymmetries, however, have turned out to be very small 
\cite{goldanskii+88,feringa+99,mills32},
and therefore their amplification is indispensable.
Frank showed theoretically that an autocatalytic reaction accompanying 
cross inhibition can lead to the amplification of 
enantiomeric excess (ee)
and to the eventual homochirality in an open system. \cite{frank53}
Following this work, numerous studies have been performed on the chiral amplification and selection in various systems. \cite{calvin69,goldanskii+88,gridnev06,sandars03,brandenburg+04}

Recently, the amplification of ee (but not homochirality)
was realized in experiments carried out by  Soai et al., 
\cite{soai+95,soai+00} and 
the temporal evolution of the chemical reaction was
shown to be explained by a second-order autocatalytic reaction
\cite{sato+01,sato+03}.
Stimulated by these works, 
we proposed that, in addition to the nonlinear autocatalytic reaction,
a recycling process induced by a back reaction gives rise to the
complete homochirality in a closed system
\cite{saito+04a}.
Subsequently several theoretical works related to this mechanism
have been done
\cite{saito+04b,saito+05a,saito+05b,saito+05c,shibata+06,
saito+07a}.

In these studies of chiral amplification, the autocatalytic reaction plays an
essential role either in open systems or in closed systems.
\cite{frank53,calvin69,goldanskii+88,gridnev06,sandars03,brandenburg+04,saito+04a,saito+04b,saito+05a,saito+05b,saito+05c,shibata+06,saito+07a}
So far, however, any autocatalytic reaction has not been found
 in the process of
polymerization, relevant for the formation of organic molecules
in life.
Granting that some pertinent
 autocatalytic reaction may well be discovered
in future, it seems worthwhile to explore possibilities of chirality selection
in non-autocatalytic way. In this paper, we demonstrate
that complete chirality selection or homochirality is possible in a 
closed system with spontaneous production together 
with recycling cross inhibition but without
autocatalytic reaction.

Our model consists of achiral substrate molecule A and two chiral
enantiomers R and S, which are produced by spontaneous productions
\begin{align}
A \rightarrow R, \qquad A \rightarrow S
\label{eq01}
\end{align}
with the same reaction rate $k_0$.
Furthermore, R and S are assumed to react back to A as
\begin{align}
R+S \rightarrow 2 A 
\label{eq02}
\end{align}
with a reaction rate $\mu_0$. We call this reaction a recycling
cross inhibition, which looks similar to Frank's cross inhibition
but differs in that whether  R and S are recycled back
in the present model
or eliminated out of the system 
in the Frank's open model.\cite{frank53}

In \S2, 
the rate equation approach shows that 
the system has a line of fixed points and chirality selection is impossible.
Since the fixed line is neutral in stability, 
the system is expected to be susceptible to the weakest perturbations  such as
fluctuation.
The rate equation, however, describes only the evolution of average quantities.
To include fluctuation effect,
one has to consider stochastic aspects of the system evolution. 
This feature can be taken into account in a stocahstic master equation approach, where the system is described by a probability distribution function.
\cite{lente04,lente05,saito+07b}
From stochastic analysis in \S3 and \S4,
  it will be shown that
the fluctuation drives the system to homochirality.
The effect of the fluctuation is attributed to the discreteness of the
microscopic process, 
the essence of which is extracted in the system size expansion 
in \S 5.
The result is summarized in \S6.

\section{Rate equation approach}

In the rate equation approach, 
the reaction
 processes, Eq.(\ref{eq01}) and Eq.(\ref{eq02}), are expressed as
\begin{align}
\frac{dr}{dt} & = k_0 a - \mu_0 rs,
\label{eq03}
\\
\frac{ds}{dt} & = k_0 a - \mu_0 rs,
\label{eq04}
\intertext{together with}
c&=a+r+s
\label{eq05}
\end{align}
where $a,r,s$ are concentrations of species A, R, S respectively
and the total concentration $c$ is assumed to be constant.
The conservation of total concentration expressed by Eq.(\ref{eq05})
implies that R and S are recycled back to A via the cross inhibition reaction
$(-\mu_0 r s)$.
The trajectories of the evolution are easily obtained as $r-s=$constant,
as shown by lines with arrows in Fig. 1.
The final states are obtained by solving $\dot r=\dot s =k_0 a - \mu_0 r s=0$
together with $a+r+s=c$, resulting the following hyperbola
\begin{align}
\Big( \frac{r}{c}+ \frac{k_0}{c\mu_0} \Big)
\Big( \frac{s}{c}+ \frac{k_0}{c\mu_0} \Big)
= \frac{k_0}{c\mu_0} \Big( 1+ \frac{k_0}{c\mu_0} \Big)
\label{eq06}
\end{align}
as is shown in Fig. 1.
There, the rate for the cross inhibition $\mu_0$ is chosen to be very large
compared to that for the spontaneous production $k_0$ as $c\mu_0 = 5k_0$,
in order to 
draw the fixed line of hyperbola clearly visible away from the diagonal boundary line $r+s=c$.
We expect, 
however, that
the cross inhibition, if it exists,
 should be a very rare process
and $c\mu_0 \ll k_0$.
The conclusion of this rate equation approach is that the enantiomeric
excess (ee) defined as
\begin{align}
\phi = \frac{r-s}{r+s} 
\label{eq07}
\end{align}
takes any value ($-1 \le \phi \le 1$), depending on initial conditions,
thus indicating no chirality selection.

%%%%%%%%%%%%%%%%%%%%%%%%%%%%%%%%%%%%%%%%%%%%%%%%%%%%%%%%%%%%%%%%%%%%%%%%%%%%%%%
\begin{figure} [htbp]
\begin{center}
\includegraphics[width=6cm,clip] {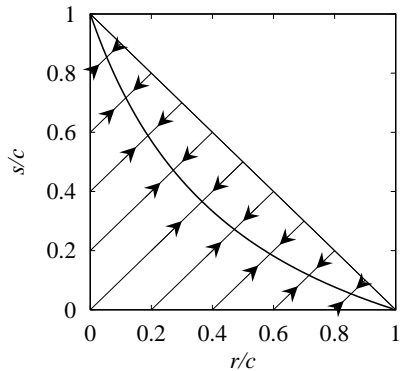}
\end{center}
\caption{Trajectory of the concentration $r$ and $s$, 
prescribed by the rate equations
(\ref{eq03}) and (\ref{eq04}),
with a fixed line of hyperbola.
The parameter is set as $c \mu_0=5k_0$ .
}
\label{fig1}
\end{figure}%
%%%%%%%%%%%%%%%%%%%%%%%%%%%%%%%%%%%%%%%%%%%%%%%%%%%%%%%%%%%%%%%%%%%%%%%%%%%%%%%%

There is, however, a subtle feature such that, along the fixed line (hyperbola),
the system is neither stable nor unstable, namely it is neutral.
Fluctuations could be decisive to destruct this neutrality and resolve the hyperbola
into a few fixed points.
In fact, a similar situation with a neutral fixed line appears 
for a closed system with a nonlinear autocatalytic process.
We have adopted a stocahstic master equation approach for this sytem,
and found chiral symmetry breaking in a previous paper. \cite{saito+07b}
Therefore, we adopt the same approach for the present system
in the following sections.

\section{Master equation approach}

For the stochastic approach, 
the relevant system is to be described in a microscopic way.
Namely, the system is
confined in a fixed volume $V$, containing species A, R, S with 
population numbers $N_A, N_R, N_S$, respectively.
Total population number of all molecules is assumed to be constant 
$N=cV$ 
since the system is closed,
 so that we have
\begin{align}
N=N_A+N_R+N_S.
\label{eq08}
\end{align}

Chemical reaction is assumed to be stochastic 
where a microscopic state specified by the population number
$\bm X=(N_A, N_R, N_S)$ varies according to a certain transition 
probabilities $W(\bm X; \bm q)$ 
of a jump $\bm q=(q_A, q_R, q_S)$
to another state $\bm X'=\bm X + \bm q$.
The probability $P(\bm X, t)$ of a state $\bm X$ at time $t$ then
evolves according to 
the master equation
\begin{align}
\frac{\partial P(\bm X, t)}{\partial t}=
\sum_{\bm q} W(\bm X- \bm q; \bm q) P(\bm X - \bm q, t)
-\sum_{\bm q} W(\bm X; \bm q) P(\bm X, t).
\label{eq10}
\end{align}
The summation by $\bm q$ is restricted by the conservation condition
Eq.(\ref{eq08}).
The transition probabilities for the present model 
is explicitly
 expressed as
\begin{align}
W(N_A, N_R, N_S; -1, +1, 0)= k_0 N_A,
\nonumber \\
W(N_A, N_R, N_S; -1, 0, +1)= k_0 N_A,
\nonumber \\
W(N_A, N_R, N_S; +2, -1, -1)= \mu N_R N_S
\end{align}
and those with other 
$\bm q$'s vanish. 
As is confirmed later, 
the coefficient for the microscopic cross inhibition $\mu$
is related to the macroscopic reaction rate $\mu_0$ as
\begin{align}
\mu= \frac{\mu_0}{V}=\frac{c\mu_0}{N}.
\label{eq12}
\end{align}
Thus the master equation in concrete form is expressed as
\begin{align}
\frac{\partial P(N_A, N_R, N_S, t)}{\partial t}=&
k_0(N_A+1)\Big\{
P(N_A+1, N_R-1, N_S,t) + P(N_A+1, N_R, N_S-1,t) \Big\}
\nonumber \\
&+\mu(N_R+1)(N_S+1) P(N_A-2, N_R+1, N_S+1,t)
\nonumber \\
&-\{2k_0 N_A+\mu N_R N_S\} P(N_A, N_R, N_S,t).
\label{eq13}
\end{align}

%%%%%%%%%%%%%%%%%%%%%
\begin{figure}[h]
\begin{center} 
\includegraphics[width=0.3\linewidth]{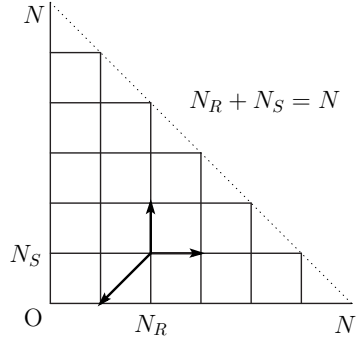}
\end{center} 
\caption{Random walk in a triangular region in $N_R-N_S$ space.
A walker at a site $(N_R, N_S)$ jumps along square edges or
along a diagonal indicated by arrows.
}
\label{fig2}
\end{figure}
%%%%%%%%%%%%%%%%%%%%%

Because of the conservation condition Eq.(\ref{eq08})
the microscopic state of the system is 
specified
 by two independent variables
$N_R$ and $N_S$ in a triangular region
\begin{align}
0 \le N_R, N_S, N_R+N_S \le N,
\label{eq14}
\end{align}
as shown in Fig. 2.
Transition probabilities $W$ 
connect
 neighboring states linked along square edges or along a diagonal,
 indicated in Fig. 2.
The system is equivalent to the directed random walk model
where a random walker jumps to the right (+1 in $N_R$ direction),
to upwards (+1 in $N_S$ direction) and to the left-down diagonal
( $-$1 in both $N_R$ and $N_S$ directions)
in the $N_R-N_S$ phase space.
One then notices easily that two homochiral states
$(N_R, N_S)=(N,0)$ and $(0,N$) are special;
there is only inflow but no outflow of the walker, namely 
the transition probabilities from the microscopic homochiral states
$(N_A, N_R, N_S)=(0, N,0)$ and $(0,0,N)$ to other states are zero,
\begin{align}
W(0,N,0; \bm q)=W(0,0,N; \bm q)=0
\label{eq15}
\end{align}
for any jump $\bm q$, 
so that, once the system enters into this homochiral states, it remains
there.
In this sense, the homochiral states are regarded as a sink or a kind of
black hole.
All other microscopic states are connected directly or indirectly 
to these two homochiral states,
and consequently the probabilities of the other states evolve into zero,
namely $P(N_A, N_R, N_S; t=\infty)=0$
for non-homochiral states.
This can be demonstrated step-by-step as follows. Because of 
Eq.(\ref{eq15}),
the master equation for the homochiral state takes the form
\begin{align}
\frac{\partial P(0, N, 0, t)}{\partial t}=&
k_0 P(1, N-1, 0,t). 
\label{eq16}
\end{align}
In the asymptotic limit where every temporal evolution has died out
$\partial P/\partial t=0$,
the asymptotic value for this neighboring state 
becomes $P(1, N-1, 0,t=\infty) =0$.
Furthermore, asymptotic limit of the master equation 
for a general state $\bm X$
becomes
\begin{align}
\sum_{\bm q} W(\bm X- \bm q; \bm q) P(\bm X - \bm q, \infty)
=\sum_{\bm q} W(\bm X; \bm q) P(\bm X, \infty)
\label{eq17}
\end{align}
so that if $P(\bm X, \infty)=0$, then all the probabilities of states
 $\bm X'$ connected directly to the state $\bm X$ 
 with positive $W$'s must have  vanishing asymptotic values 
 $P(\bm X', \infty)=0$.
 Since every state is interconnected, this completes the proof.
 
%%%%%%%%%%%%%%%%%%%%%%%%%%%%%%%%%%%%%%%%%%%%%%%%%%%%%%%%%%%%%%%%%%%%%%%%%%%%%
\begin{figure} [htbp]
\begin{center}
\includegraphics[width=0.32\linewidth] {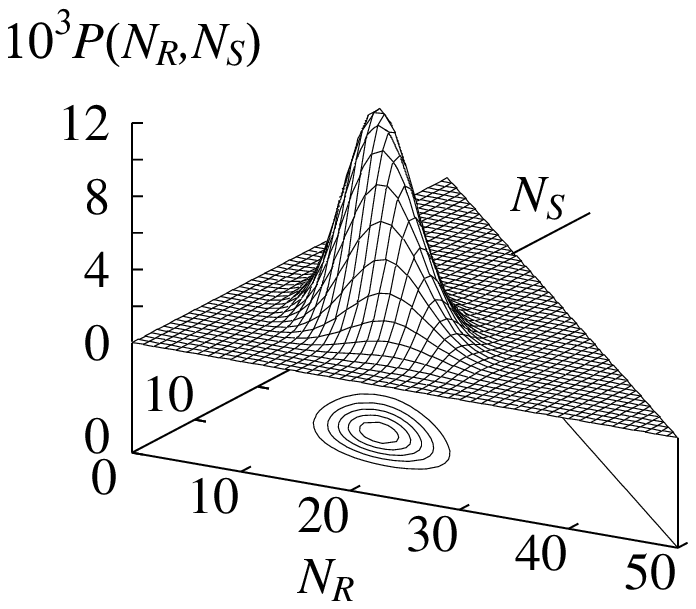}
\includegraphics[width=0.32\linewidth] {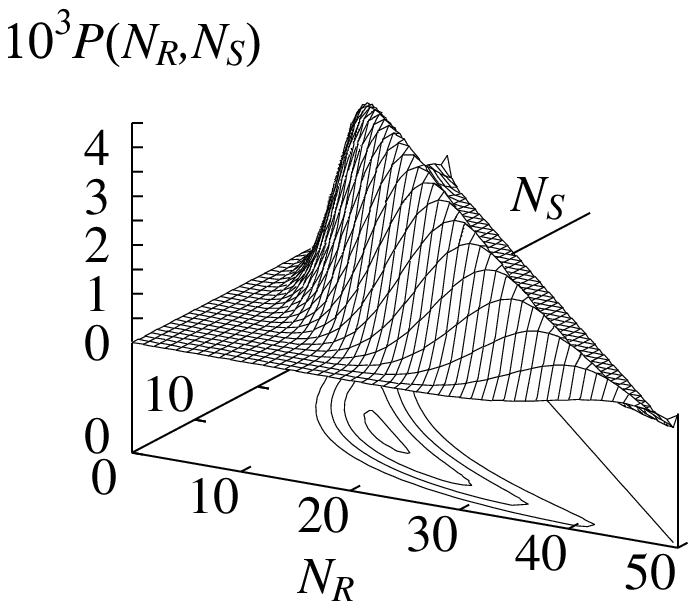}
\includegraphics[width=0.32\linewidth] {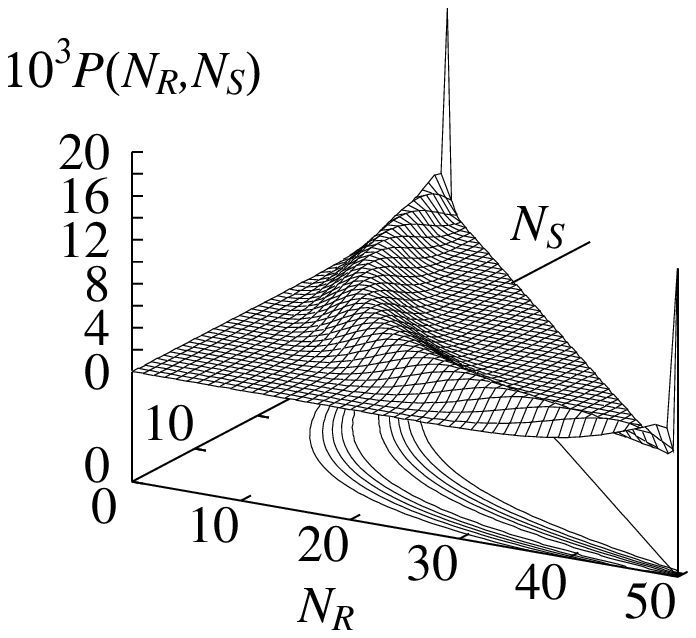}\\
(a) \hspace{4cm} (b) \hspace{4cm} (c)
\caption{
Time evolution of the probability distribution 
with trajectory of the probability contour at the basement,
obtained by numerically integrating
 the master equation (\ref{eq10}) for $\mu=k_0/10$.
 The system starts from $P(N,0,0)=1$ and depicted times are at
(a)$k_0t=1$, (b)$k_0t=9$  and (c)$k_0t=19$.
The total population number is $N=50$. }
\label{fig3}
\end{center}
\end{figure}%
%%%%%%%%%%%%%%%%%%%%%%%%%%%%%%%%%%%%%%%%%%%%%%%%%%%%%%%%%%%%%%%%%%%%%%%%%%%%%%
This conclusion is confirmed numerically by solving the time development
of the master equation Eq.(\ref{eq13}) 
by using the Runge-Kutta method of the
fourth order.\cite{press+92} 
Starting from a completely achiral initial condition
with $N_A=N$ and no chiral ingredients $N_R=N_S=0$, 
the probability distribution, started from the initial delta-peak
$P(N_A,N_R, N_S, t=0)=\delta_{N_A,N} \delta_{N_R,0}\delta_{N_S,0}$,
remains symmetric at any time, as shown in Fig. 3.
There, the redundant variable $N_A=N-N_R-N_S$ is suppressed, and
the probability distribution is shown in $N_R$ and $N_S$ phase space.
The probability contour is also shown at the basement.
In this numerical calculation, the total population number is set as 
$N=50$ due to the limitation of the calculationa capacity and $\mu=k_0/10$ 
(or $c \mu_0=5k_0$ as in Fig. 1) for the sake of visibility of distribution.
In the early stage at a time $k_0t=1$ in Fig.3(a), 
the probability distribution has an approximately 
Gaussian shape with a central peak at the racemic
fixed point
$N_R=N_S=(k_0/\mu)( \sqrt{N\mu/k_0+1}-1)\approx 14.5$
 in agreement with results obtained by the rate eqautions.
At the intermediate time $k_0t=9$, the probability 
distibution spreads along the 
hyperbola which is the fixed line found 
in the rate equation approach, 
as shown in Fig.3(b).
In the late stage, the probability disitribution develops sharp peaks at
the two end points corresponding to the two homochiral states (Fig. 3(c)).

By the use of this probability distribution, 
an expectation value of any function 
$f(N_A, N_R, N_S)$ at a time $t$ is easily calculated as
\begin{align}
\langle f(N_A, N_R, N_S) \rangle_t=\sum_{N_A, N_R, N_S} 
f(N_A, N_R, N_S) P(N_A, N_R, N_S, t)
\label{eq18}
\end{align}
In Fig. 4, 
the time development of the expectation value of the
population number of R species $\langle N_R \rangle_t $
is shown.
In the early stage in Fig. 4(a), the average $\langle N_R \rangle_t $
increases sharply to the racemic fixed point value 14.5 
predicted by 
the rate equation within a time scale of $k_0t \approx 1$.
This development is well described by the rate 
equations Eq.(\ref{eq03}) and Eq.(\ref{eq04}) together with  Eq.(\ref{eq05}),
 as is shown by a dashed curve in Fig. 4(a).
The results by the rate equations completely agrees with 
the evolution of average value $\langle N_R \rangle_t $ 
until $k_0t \approx 1$.
Then, 
the rate equations predict the saturations at the racemic value,
but the numerical simulation of the master equations indicates a slow increase.
The average $\langle N_R \rangle_t $ approaches ultimately to the 
value $N/2$, corresponding to the double peak profile of 
the final probability distribution at the two homochiral states.
The slow approach is expressed well by the following form
\begin{align}
\langle N_R \rangle_t =\frac{N}{2} - A e^{-t/\tau} .
\end{align}
This exponential behavior is evident in Fig. 4(b) where the 
logarithmic difference
$\ln [{N}/{2} -\langle N_R \rangle_t]$ is plotted versus $k_0t$.
The fitting gives values $A=10$ and $1/k_0\tau=0.0163$.

%%%%%%%%%%%%%%%%%%%%%%%%%%%%%%%%%%%%%%%%%%%%%%%%%%%%%%%%%%%
\begin{figure}[htbp]
\begin{center}
\includegraphics[width=6cm,clip] {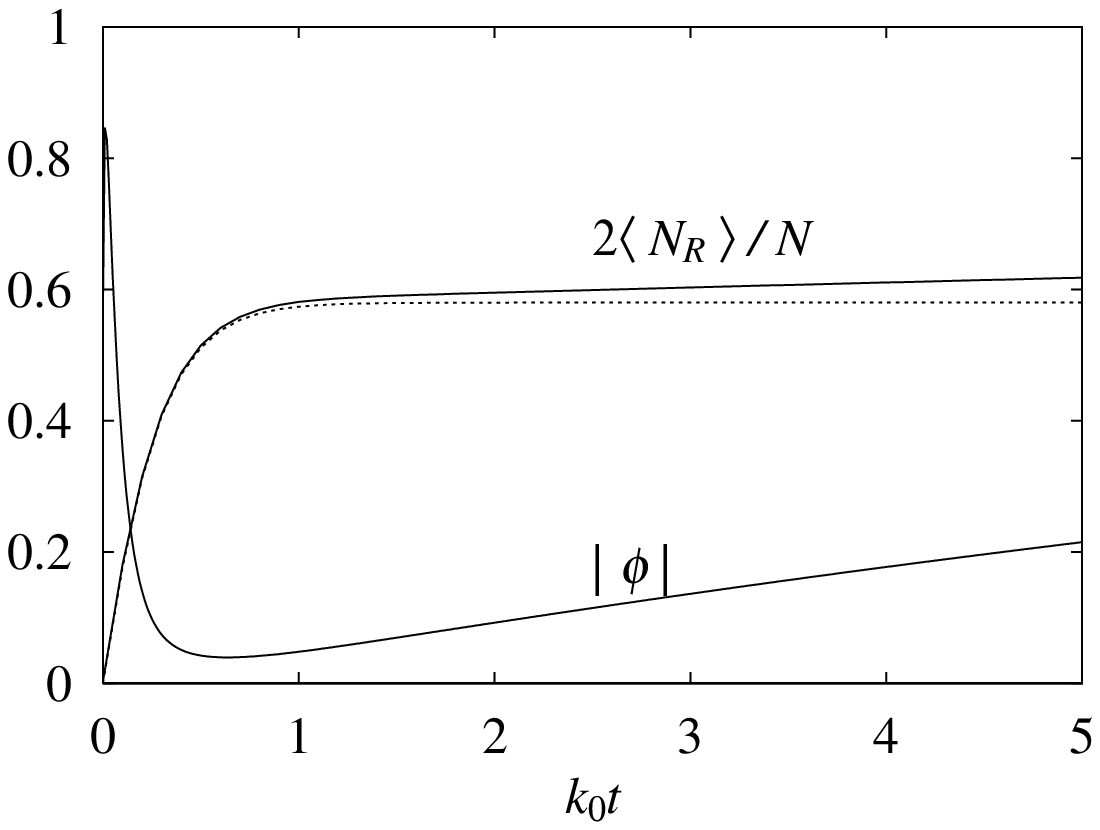}
\includegraphics[width=6cm,clip] {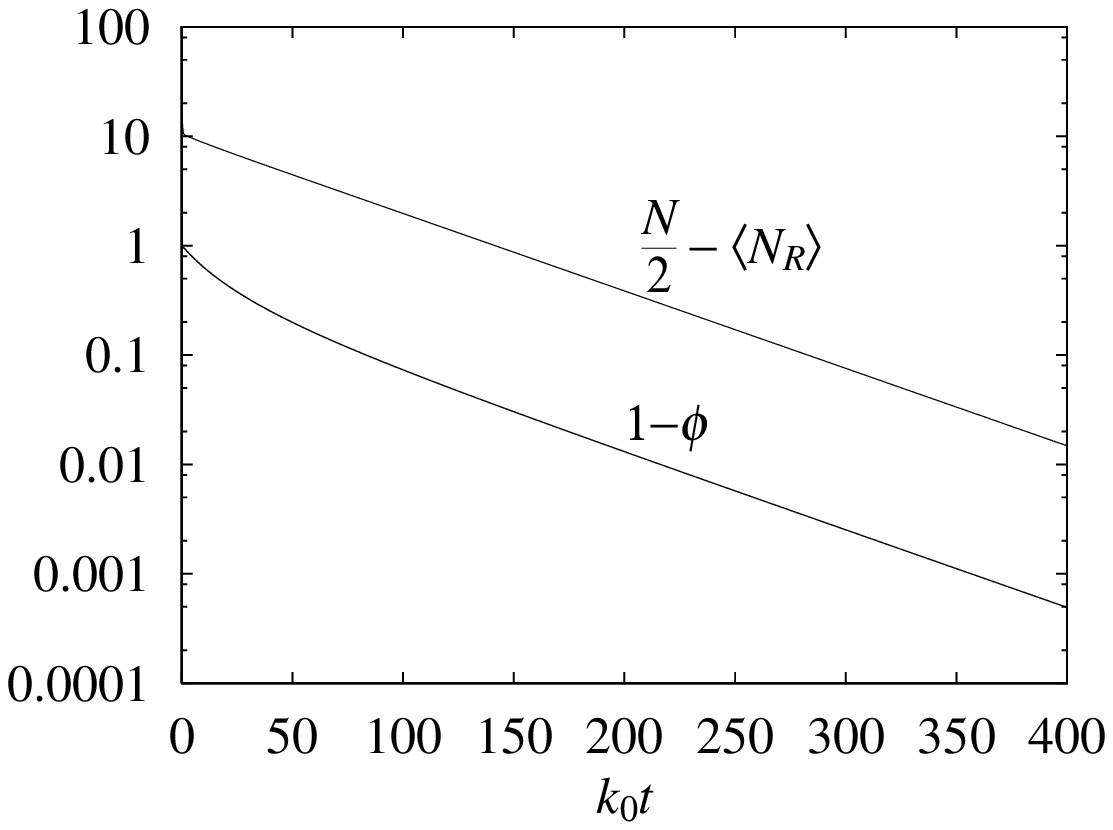}\\
(a) \hspace{6cm} (b)\\
%\includegraphics[width=6cm,clip] {fig4c.eps}
%\includegraphics[width=6cm,clip] {fig4ab.eps}\\
%(c) \hspace{6cm} (d)\\
\caption{
Time development of the average population $\langle N_R \rangle_t $ and 
 the absolute value of the ee order parameter $|\phi|$.
(a) In an early stage, $k_0 t \le 1$, 
$\langle N_R \rangle_t $ agrees with the evolution by the rate equation, 
which
 is shown by a dashed curve.
(b) Exponential relaxation 
in
 the asymptotic region.}
\label{fig4}
\end{center}
\end{figure}
%%%%%%%%%%%%%%%%%%%%%%%%%%%%%%%%%%%%%%%%%%%%%%%%%%%%%%%%%
\begin{figure}[htbp]
\begin{center}
\includegraphics[width=6cm,clip] {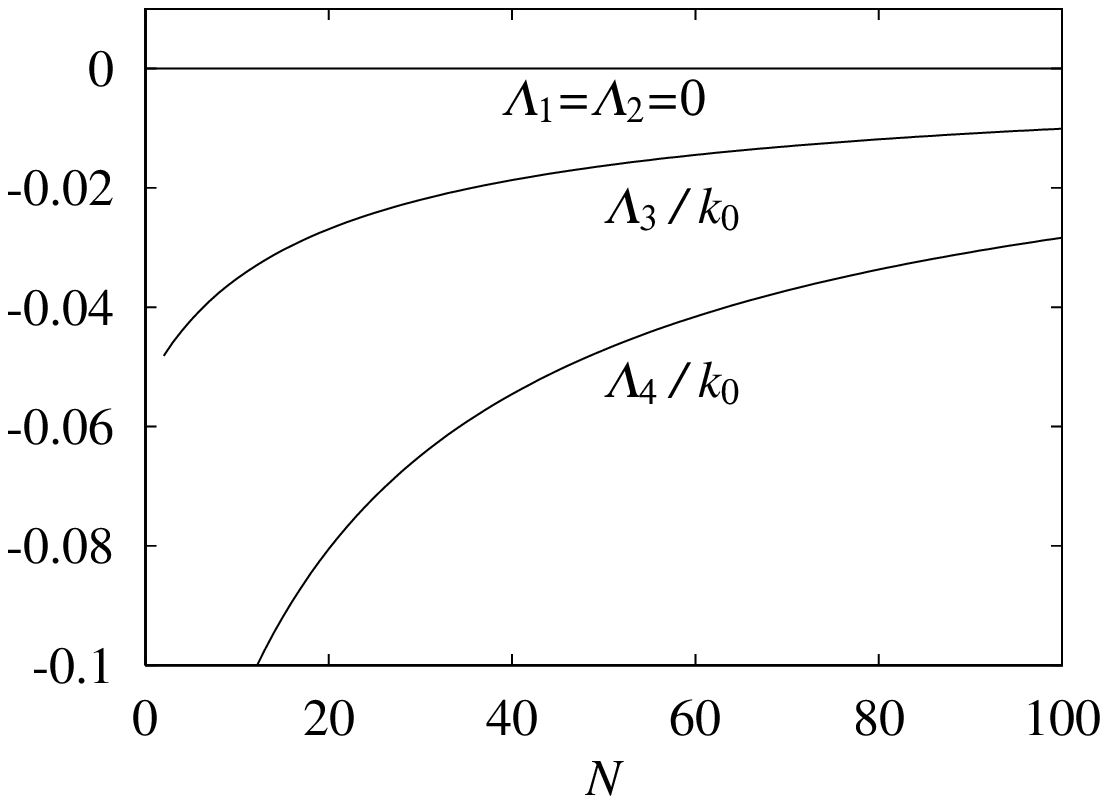}
\includegraphics[width=6cm,clip] {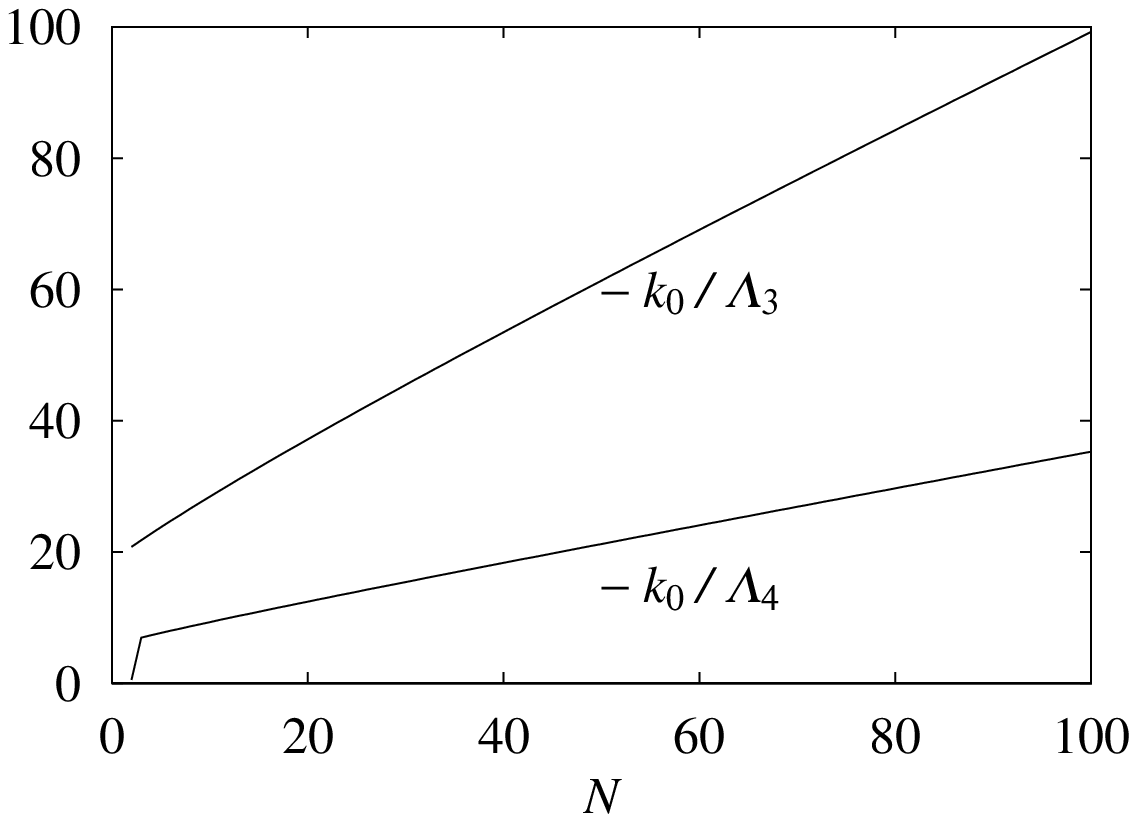}\\
(a) \hspace{6cm} (b)\\
\includegraphics[width=6cm,clip] {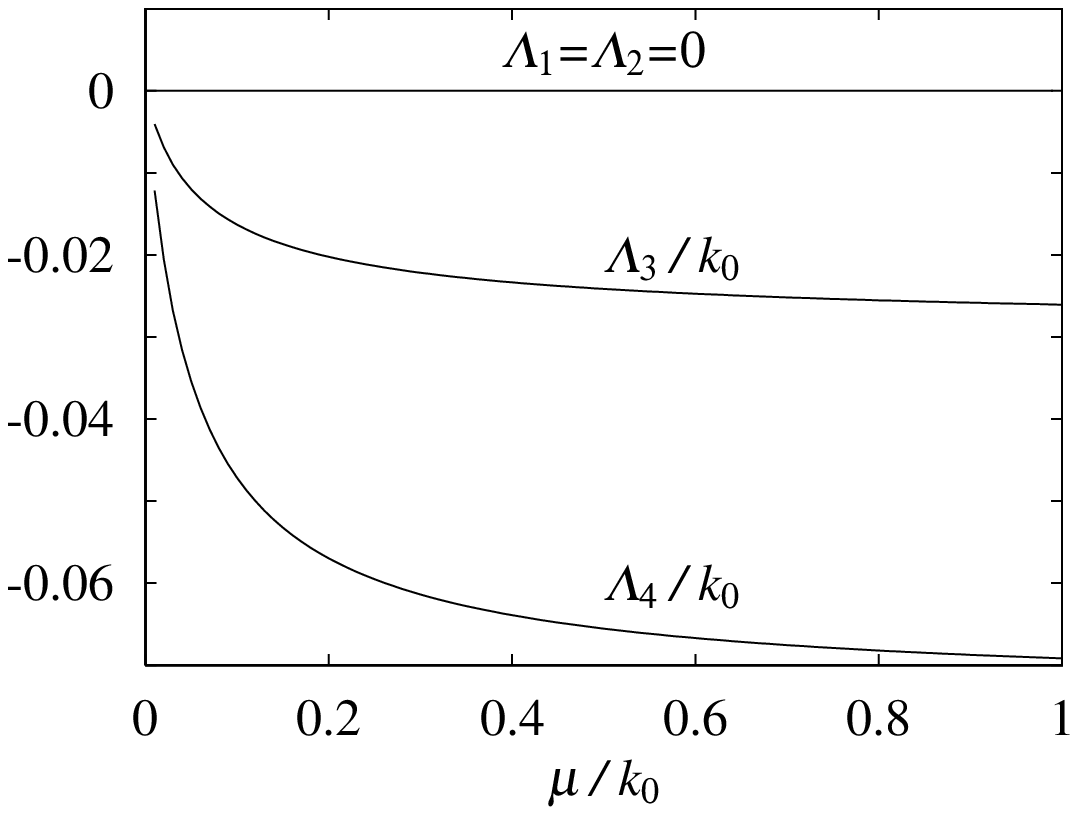}
\includegraphics[width=6cm,clip] {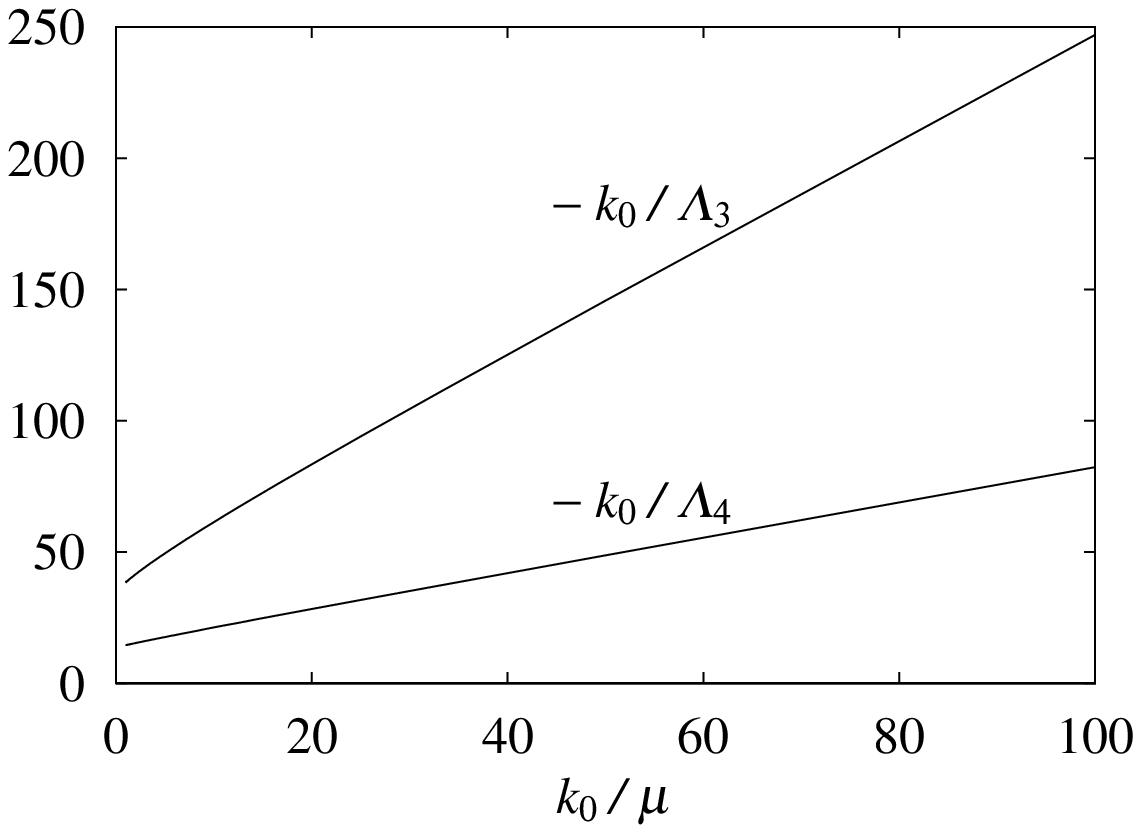}\\
(c) \hspace{6cm} (d)\\
\caption{
 A few largest eigenvalues $\Lambda_i$
of the time-evolution matrix 
plotted (a) versus total population $N$ at a fixed 
cross inhibition, $\mu/k_0 =0.1 $, 
and (c) versus $\mu/k_0$ at a fixed $N=50$. 
By plotting the inverse of eigenvalues in (b) and (d), one finds
$k_0/|\Lambda_i|$ is linear in $N$ and also in 
$k_0/\mu$, asymptotically.
}
\end{center}
\label{fig5}
\end{figure}
%%%%%%%%%%%%%%%%%%%%%%%%%%%%%%%%%%%%%%%%%%%%%%%%%%%%%%%%%

The enantiomeric excess (ee) corresponds to an order parameter 
in a phase transition in  standard statistical mechanics.
Adopting an analogous definition of an order parameter in numerical simulations
for magnetic phase transitions, we define the absolute value
of ee as
\begin{align}
|\phi(t)| = \Big\langle \Big(\frac{N_R-N_S}{N_R+N_S}\Big)^2 
\Big\rangle^{1/2}_t.
\label{eq19}
\end{align}
The time development of an ee order parameter $|\phi|$
is also shown in Fig.4(a) and (b). In the very early stage 
$k_0 t \le 0.2$ in Fig. 4(a)
$|\phi|$ is very large because the probability has a large amplitude close to the edges $N_R=0$ or $N_S=0$.
As time evolves, the numbers of R and S molecules increase 
and the peak position of the probability distribution
leaves from both edges
(Fig. 3(a)),
 and thus the ee value $|\phi|$ drops sharply.
After the peak of the probability distribution
reaches the racemic fixed point where $|\phi| \approx 0$, 
the probability spreads along the fixed line 
(Fig. 3(b)),
and thus $|\phi|$ value increases steadily to a final value, unity, 
of homochirality.
By plotting the logarithm of the difference, $\ln[1-|\phi(t)|]$,
as a function of $k_0t$, as in Fig. 4(b), 
one again obtains the exponential relaxation with
the same exponent $1/k_0\tau \approx 0.0162$.

\section{Eigenvalue Analysis}

The asymptotic relaxation of the average value and the ee order parameter
turned out to be exponential with the same characteristic time,
$k_0\tau \approx 60$. We consider this problem in terms of eigenvalues
of 
evolution matrix of
the master equation. Since master equation is a linear equation
for the probability distribution, the time evolution is written by 
using
the evolution matrix $\bm M$ 
\begin{align}
\frac{dP(\bm X, t)}{dt}= \sum_{\bm X'}(\bm X|\bm M|\bm X')P(\bm X',t),
\label{eq21}
\end{align}
where matrix elements of $\bm M$ are
related to the transition probabilities $W$ as
\begin{align}
(\bm X|\bm M|\bm X')=
\begin{cases}
W(\bm X'; \bm X-\bm X') & \mbox{for} \quad \bm X' \ne \bm X,\\
-\sum_{\bm q} W(\bm X; \bm q) & \mbox{for} \quad \bm X'=\bm X.\end{cases}
\label{eq22}
\end{align}
By the use of eigenfunctions $\Psi_i$ and eigenvalues $\Lambda_i$ of the
matrix $\bm M$ as
\begin{align}
\bm M \Psi_i = \Lambda_i \Psi_i ,
\label{eq23}
\end{align}
the time development of the probability distribution is expressed as a series
\begin{align}
P(\bm X,t) = \sum_{i=1}^{\infty} a_i e^{\Lambda_i t} \Psi_i .
\label{eq23}
\end{align}
Since the probability distribution satisfies the conservation as
$\sum_{\bm X} P(\bm X,t)=1$, all the eigenvalues must 
be non-positive.
In the previous section, we have proven that two homochiral states
corresponds to the final states, so that there are two
degenerate zero eigenvalues $\Lambda_1=\Lambda_2=0$
with eigenstates 
$\Psi_1=\delta_{N_A,0}\delta_{N_R,N}\delta_{N_S,0}$ and
$\Psi_2=\delta_{N_A,0}\delta_{N_R,0}\delta_{N_S,N}$,
or their linear combinations.
The asymptotic temporal evolution is governed then by the third largest
eigenvalue $\Lambda_3$. 

One can calculate eigenvalues of the matrix
$\bm M$ numerically by using the subroutine "dgeev" in LAPACK.
A few largest eigenvalues are shown in Fig. 5 as a function of
the total population number $N$ for a fixed cross inhibition $\mu=0.1k_0$
(Figs. 5(a) and (b)),
or as a function of $\mu/k_0$ for a fixed $N=50$
(Figs. 5(c) and (d)).
For $\mu=0.1k_0$ and $N=50$, the third largest eigenvalue 
has a value
$\Lambda_3/k_0
=-0.0163$, in good agreement with the exponent $1/k_0\tau
=0.0163$ obtained  in the previous
section from the asymptotic final relaxation of the average $\langle N_R \rangle_t $ and the ee values $|\phi(t)|$.

By plotting the inverse of eigenvalues $-k_0/\Lambda_i$
 as in Figs. 5(b) and (d), one notices that it is linear 
 in the total population number $N$ and  the inverse strength of the
 cross inhibition $k_0/\mu$, asymptotically.
Therefore, for a very large system with a small cross inhibition,
it should take a long time of the order  $N/\mu$ 
before the homochirality becomes observable.

\section{System size expansion}

For our model with spontaneous production of chiral species
with recycling cross inhibition, the rate equation tells us
no chirality selection whereas the stochastic master equation
insists the final configuration be homochiral. The totally
different conclusions are ascribed to the fluctuation due to
the discreteness of microscopic processes.
The fluctuation effect associated to the system size 
is qualitatively analyzed by the
system size 
(precisely said, the inverse system size)
expansion  analysis of the master equation, developed
by R. Kubo et al\cite{kubo+73,saito+76}.

In the master Eq.(\ref{eq10}), the probability density 
$P(\bm X, t)$ of a microscopic state $\bm X$ is connected to 
another state $\bm X+\bm q$ which differs  with a jump 
$\bm q$ of order unity by a transition
probability $W(\bm X; \bm q)$.
The rate $W$ is of macroscopic order of the system size
$N$ or the volume $V$ as 
\begin{align}
W(\bm X; \bm q)=V w(\bm x; \bm q)
\end{align}
where $\bm x$ is the density variable $\bm x=\bm X/V$ 
of order unity:
\begin{align}
\bm x = (a, r, s) =\Big(\frac{N_A}{V}, 
\frac{N_R}{V}, \frac{N_S}{V} \Big) .
\label{eq25}
\end{align}
Then, the probability is assumed to take the form
$P(\bm X, t)= \exp [V \chi(\bm x,t)]$, and
time evolutions of the leading order contributions of 
the average density $\langle \bm x \rangle_t$ 
and the correlation functions 
\begin{align}
\sigma_{ij}(t)=V \langle (x_i-\langle x_i \rangle_t)
(x_j-\langle x_j \rangle_t) \rangle
\end{align}
are shown to be determined by moments 
\begin{align}
c_{ij \cdots }(\bm x)=\sum_{\bm q} q_i q_j \cdots~ w(\bm x; \bm q)
\end{align}
of the transition probability $w$ as
\begin{align}
\frac{d }{dt} \langle x_i \rangle_t& = c_{i}(
\langle \bm x \rangle_t)
\nonumber \\
\frac{d }{dt} \sigma_{ij}(t)& = 
\sum_k \Big( \frac{\partial c_{i}}{\partial \langle x_k \rangle_t} \sigma_{kj}(t)+
\sigma_{ik}(t) \frac{\partial c_{j}}{\partial \langle x_k \rangle_t} \Big) +
c_{ij}(\langle \bm x \rangle_t) .
\end{align}
The evolution equations for the higher order corrections can also be 
derived.\cite{kubo+73}
If the system is normal, the fluctuation correlations $\sigma_{ij}$
remains of order unity. On the other hand, if the system is unstable,
as in the case of phase transitions with critical behaviors,
at least one of the fluctuations $\sigma_{ij}$ is enhanced 
 to the order of the system size.

In the present model, first order moments are
\begin{align}
c_r=c_s=k_0a-\mu_0rs=k_0(c-r-s)-\mu_0rs
\end{align}
and the second order moments are
\begin{align}
c_{rr}=c_{ss}=k_0a+\mu_0rs=k_0(c-r-s)+\mu_0rs , \quad c_{rs}=\mu_0 rs  ,
\end{align}
where $\mu_0=V \mu$ as defined previously in Eq.(\ref{eq12}).
Thus, 
the lowest order of 
the average concentrations $\langle r \rangle_t$ and 
$\langle s \rangle_t$ satisfy the rate equations Eq.(\ref{eq03}) and Eq.(\ref{eq04}). 
For simplicity, we describe these lowest order average values as $r$ and $s$,
hereafter.
The correlation functions follow the evolution
\begin{align}
&\frac{d}{dt} \sigma_{rr}(t)=-2(k_0+\mu_0s)\sigma_{rr}-2(k_0+\mu_0r)\sigma_{rs}
+k_0(c-r-s)+\mu_0rs
\nonumber \\
&\frac{d}{dt} \sigma_{rs}(t)=-(k_0+\mu_0s)\sigma_{rr}-(2k_0+\mu_0r+\mu_0s)\sigma_{rs}-(k_0+\mu_0 r)\sigma_{ss} + \mu_0 rs
\nonumber \\
&\frac{d}{dt} \sigma_{ss}(t)=-2(k_0+\mu_0r)\sigma_{ss}-2(k_0+\mu_0s)\sigma_{rs}
+k_0(c-r-s)+\mu_0rs
\end{align}
To detect the chiral symmetry breaking, it is more convenient to use the following 
 symmetric and asymmetric variables
\begin{align}
x_+=r+s, \quad x_-=r-s.
\end{align}
Their averages and correlation functions evolve as
\begin{align}
\frac{d}{dt} x_+ &=2k_0(c-  x_+)-\frac{1}{2}\mu_0( x_+^2-  x_-^2), \quad
\frac{d}{dt}    x_-=0,
\nonumber \\
\frac{d}{dt} \sigma_{++}&=-2(2k_0+\mu_0  x_+) \sigma_{++}+
2k_0(c-  x_+)+\mu_0  (x_+ ^2-x_-^2),
\nonumber \\
\frac{d}{dt} \sigma_{+-}&=-(2k_0+\mu_0  x_+) \sigma_{+-}+\mu_0x_-\sigma_{--},
\quad
\frac{d}{dt} \sigma_{--}=2k_0(c-  x_+).
\label{eq33}
\end{align}
By starting from the completely achiral initial condition, the system remains
racemic with average value
 $ x_- =0$,
and  $ x_+ $ approaches to the racemic fixed point value
$(2k_0/\mu_0)(\sqrt{c\mu_0/k_0+1}-1)$.
The time the system takes to reach the racemic fixed point is of order $1/k_0$. 

The fluctuation of the asymmetry variable $\sigma_{--}$, however, 
increases indefinitely in this approximation.
In fact, when the average value of $x_+$ takes the racemic point value,
the fluctuation increases linearly in time
with a positive velocity; 
\begin{align}
\frac{\dot \sigma_{--}}{c}=\frac12 \frac{\mu_0}{c} x_+^2
=\frac12 N\mu\left(\frac{x_+}{c}\right)^2>0.
\label{eq34}
\end{align}
This increase of the fluctuation indicates 
that the discreteness in the microscopic process of chemical reaction 
evokes an intrinsic instability of the racemic fixed point
 (and in general, every points on the fixed line) 
to macroscopic level.

%%%%%%%%%%%%%%%%%%%%%%%%%%%%%%%%%%%%%%%%%%%%%%%%%% 
\begin{figure} [htbp]
\begin{center}
\includegraphics[width=0.4\linewidth] {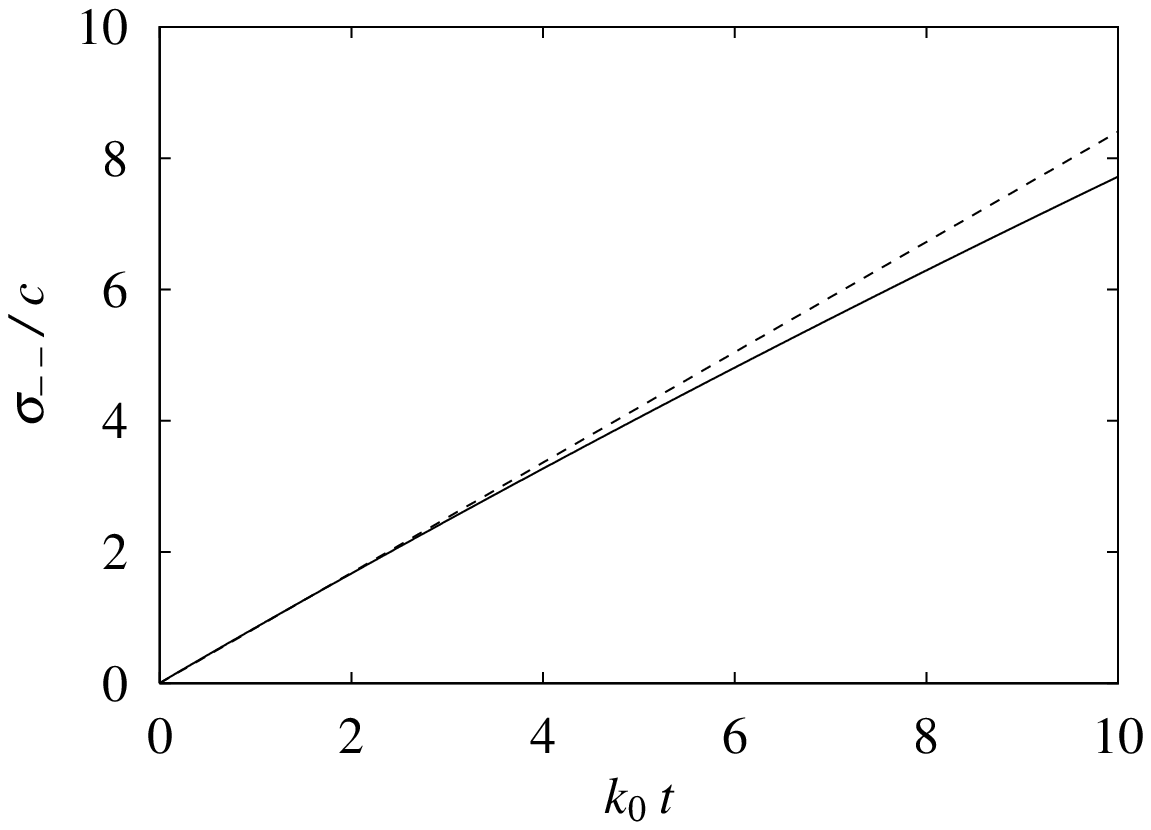}
\hspace{0.5cm}
\includegraphics[width=0.4\linewidth] {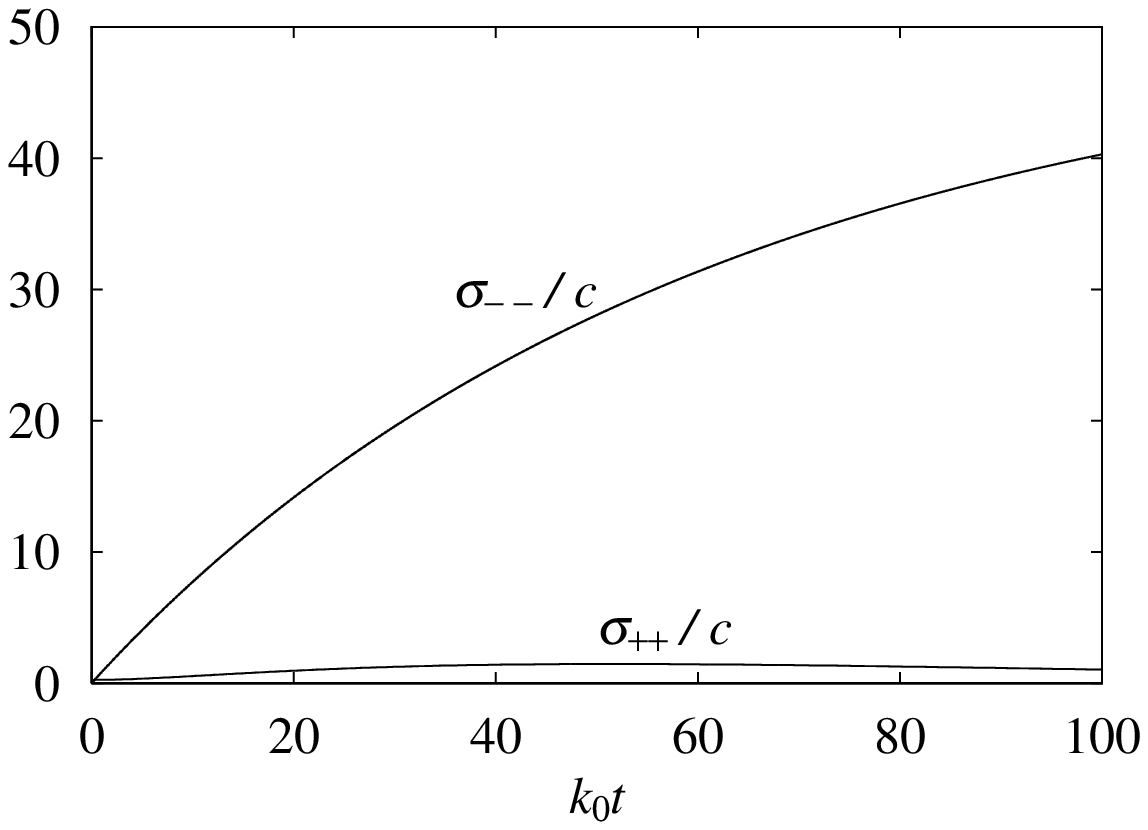}\\
(a) \hspace{6cm}  (b)
\caption{
Fluctuation correlation functions obtained 
by numerically time integrating the master equation 
(a) at an early stage, and (b) a whole time range.
Parameter is set as $\mu=k_0/10$, or $c\mu_0=5k_0$.
A dotted line in (a) represents the result of system size expansion.
}
\label{fig6}
\end{center}
\end{figure}%
%%%%%%%%%%%%%%%%%%%%%%%%%%%%%%%%%%%%%%%%%%%%%

In the numerical simulation the fluctuation of asymmetric variable 
$\sigma_{--}(t)/c$ actually increases in time, as shown in Fig. 6.
The initial increase shown by a continuous curve in Fig. 6(a)
 is linear in time $k_0t$, as expected from the 
system size expansion analysis, shown by a dashed line.
As time passes, the system size analysis fails, since the
average value $x_+$ deviates from the racemic value in the actual simulation.
One then has to consider higher order contributions to the average of $x_+$.
Departure from the racemic state becomes visible in a macrosopic sense 
when the fluctuation expressed by $\sigma_{--}/c$ reaches the order $N$. 
The time for this to appear is of order $1/\mu$, as is obtained from Eq.(\ref{eq34}) and by assuming $x_+ \sim c$.

Asymptotically, the double peak structure developes in the probability
distribution. The time scale for this to fully develop is governed by 
the largest nonzero eigenvalue $\Lambda_3$ so that the homochirality is 
realized in the time of order of 
$1/|\Lambda_3|\approx N/\mu$,
as is described in the previous section.

\section{Conclusion and discussion}

It seems to be a general consensus that an autocatalytic production process, either linear or nonlinear, is indispensable for the realization of homochirality.
 In the present work, we proposed a new mechanism by using a simple model
 to demonstrated that the homochirality can be realized without any autocatalytic production process. Our model consists of a spontaeous production together 
 with a recycling cross inhibition in a closed system. As was shown, the 
 rate equations for this system predict no chiral symmetry breaking, but 
 the stochastic master equation predicts complete homochirality. 
 This is because the fluctuation induced by the discreteness  of population numbers of participating molecules plays essential roles. This fluctuation conspires with the recyling cross inhibition to realize the homochirality. 
 
 If this fluctuation mechanism could explain the homochirality in life, then 
 this is what Pearson suggested long time ago.\cite{peason898} 
 However, the necessary time for the homochirality to set 
 in due to the fluctuation is very large , as it is proportional to the total number of the relevant 
 molecules ($\approx N/\mu$), which is of macroscopic size. Taking this feature into account, 
 we can conceive a new senario for the homochirality in macroscopic scale 
 as follows. At first, in some small closed corner,
such as in a region enclosed by a vesicle, 
 the fluctuation induced 
 homochirality is realized in a very long time 
 with respect to
  the time scale of laboratory experiments,
  but not so long with respect to geological time scale.
  It may be conceivable that in this period
some unknown autocatalytic reaction, the effect of which is too small to be detected in the laboratory experiment
so far,
  begins to operate and generates the large scale realization of the homochiralty from the homochiral seeds produced by the fluctuation
  mechanism proposed here.

Even though reaction systems with a recycling cross inhibition are
not yet found,
we hope that a simple system with only a recycling cross inhibition
might be found in near future, and the establishment of
homochirality be checked.
 
 \acknowledgement

Y. S. acknowledges support by a Grant-in-Aid for Scientific Research (No. 19540410) from the Japan Society for the Promotion of Science.

%%%%%%%%%%%%%%%%%%%%%%%%%%%%%%%%%%%%%%%%%%%%%%%%%%%%%%%5

\end{document}